\documentclass{llncs}[11pt]
\usepackage{ amsmath, amsfonts, amssymb}

\newcommand{\depth}{\mathrm{depth}}
\newcommand{\size}{\mathrm{size}}

\newcommand{\Upar}[1]{U^{(#1)}_{\textup{parity}}}
\newcommand{\Ufan}[1]{U^{(#1)}_{\textup{fan-out}}}

\newcommand{\RP}{\mathrm{RP}}
\newcommand{\QNC}{\mathrm{QNC}}
\newcommand{\QNCwf}{\mathrm{QNC}_{\mathrm{f}}}
\newcommand{\UQNCwf}{\mathrm{UQNC}_{\mathrm{f}}}
\newcommand{\IQNCwf}{\mathrm{IQNC}_{\mathrm{f}}}
\newcommand{\OQNCwf}{\mathrm{OQNC}_{\mathrm{f}}}

\newcommand{\BQNCwf}{\mathrm{BQNC}_{\mathrm{f}}}
\newcommand{\BQNC}{\mathrm{BQNC}}

\newcommand{\MNC}{\mathrm{QMNC}}
\newcommand{\BMNC}{\mathrm{BQMNC}}

\def\ket#1{{|}#1\rangle}

\begin{document}

\title{Computational depth complexity of measurement-based quantum computation}
\author{Dan Browne\inst{1}, Elham Kashefi\inst{2}, Simon Perdrix\inst{2,3}}
\institute{Department of Physics and Astronomy, University College London, UK.\\
    \email{d.browne@ucl.ac.uk}
\and
Laboratory for Foundations of Computer Science, University of Edinburgh, UK\\
\email{ekashefi@inf.ed.ac.uk}
\and
Laboratoire Preuves, Programmes et Syst\`emes, Universit\'e Paris Diderot, France\\
\email{simon.perdrix@pps.jussieu.fr}
}
\date{}
\maketitle

\begin{abstract}
We prove that one-way quantum computations have the same computational power as quantum circuits with unbounded fan-out. It demonstrates that the one-way model is not only one of the most promising models of physical realisation, but also a very powerful model of quantum computation. It confirms and completes previous results which have pointed out, for some specific problems, a depth separation between the one-way model and the quantum circuit model. 
Since one-way model has the same computational power as unbounded quantum fan-out circuits, the quantum Fourier transform can be approximated in constant depth in the one-way model, and thus the factorisation can be done by a polytime probabilistic classical algorithm which has access to a constant-depth one-way quantum computer. 
The extra power of the one-way model,  comparing with the quantum circuit model, comes from its classical-quantum hybrid  nature. We show that this extra power is reduced to the capability to perform unbounded classical parity gates in constant depth.

 \end{abstract}

\section{Introduction}

The one-way quantum computational model, proposed by Raussendorf and Brie\-gel \cite{RB00}, is remarkable in many aspects. It represents an approach to quantum computation very different to more conventional ``circuit-based'' approaches which were derived in close analogy to classical logic circuits. In the one-way model, computation proceeds by the generation of a particular entangled multi-qubit state - a cluster state - followed by the adaptive measurement of individual qubits. The choice of basis for the measurements, and their adaptive dependency encodes the computation. 

The dependancy of the bases upon the outcome of previous measurements is a necessary part of the model. It compensates for the inherent randomness of the outcome of individual measurements allowing deterministic computation. Measurements which are not directly or indirectly dependent upon each other can be performed simultaneously. Thus the one-way model offers a radically different approach to the parallelisation of computations.

Broadbent and Kashefi \cite{BK07} have pointed out a depth separation between the quantum circuit model and the one-way model. Indeed, there is a constant depth one-way quantum computation for implementing the parity gate, whereas there is no poly-size constant-depth circuit for this gate. Moreover they have proved that for any problem, the depth-separation between quantum circuits and one-way model is at most logarithmic. However, the exact power of the one-way quantum computation remained unknown.

In this paper, we mainly prove 
 that the computational power of the one-way model is equivalent, up to classical side-processing, to the quantum circuit model augmented with unbounded fanout gates. This model, first explored by H\o yer and Spalek \cite{HS05} is a computational model which allows any two commuting gates to be performed simultaneously. The unbounded fanout model is surprisingly powerful, for example, the quantum component of Shor's algorithm reduces to constant depth in this model. Our results imply that the one-way model shares this power, provided the depth of the classical parity computations which make up the dependency calculations is neglected.

In section \ref{sec:meascal}, we review the Measurement Calculus \cite{DKP07}, a formal framework in which the one-way model can be succinctly represented. Section \ref{sec:mp-qc} is dedicated to the comparison of the computational power of the one-way model and the quantum circuit model. In section \ref{sec:fan-out} we present the unbounded fanout circuit model and prove the main result of this paper, its equivalence with the one-way model. In section \ref{sec:gen}, we show that reasonable classically controlled models can be efficiently implemented in the one-way model, answering an open question stated in \cite{KOBAA09}. In section \ref{sec:app}, we discuss some applications of these results and in section \ref{sec:assumption} we discuss the assumptions underlying each of these models.

\section{Measurement calculus}\label{sec:meascal}

One-way quantum computations can be rigourously described in the measurement calculus formalism \cite{DKP07}. A term of measurement calculus, called \emph{measurement pattern} and playing the role of a circuit in the quantum circuit model, is a quadruplet $(V,I,O, A)$. $V$ is a finite register of qubits, $I, O\subseteq V$ are subresgisters denoting respectively the inputs and output qubits. The non input qubits ($V\setminus I$) are initialised in state $\ket +=\frac1{\sqrt 2}(\ket 0+\ket 1)$. $A$ is a sequence of quantum operations, called commands. They are three kinds of commands: 
\begin{itemize}
\item  $E_{i,j}$ is an entangling operation, which is nothing but the controlled-Z $\Lambda Z$ unitary gate on qubits $i,j\in V$ where $\Lambda Z$ is definied as follows: $$\Lambda Z = \left( \begin{array}{cccc}1&0&0&0\\0&1&0&0\\0&0&1&0\\0&0&0&-1\end{array}\right)$$
\item $M_i^\alpha$ is a measurement of the qubit $i$ in the basis $\{\ket {+_\alpha}, \ket{-_\alpha}\}$:$$\ket{\pm_\alpha}=\frac1{\sqrt2}(\ket 0\pm e^{i\alpha}\ket 1)=\frac1{\sqrt2}\left(\begin{array}{c}1\\\pm e^{i\alpha}\end{array}\right)$$ This measurement produces a classical outcomes $s_i\in \{0,1\}$.
\item $X_i^{\sigma}$ and $Z_i^{\sigma}$ are Pauli corrections or \emph{dependant corrections}, where $s$ is a finite sum modulo $2$ of classical outcomes $s_j$'s. $X_i^{s_j+\ldots +s_k}$ is applied on qubit $i$ and depends on the sum of $s_j,\ldots , s_k$ as follows: $X^{0}=I$ and $X^1=X$. $$X=\left(\begin{array}{cc}0&1\\1&0\end{array}\right)~~~Z=\left(\begin{array}{cc}1&0\\0&-1\end{array}\right)$$
\end{itemize}

Measurements are supposed to be destructive (a measured qubit cannot be reused anymore), as a consequence a measurement pattern is well-formed if no command is applied on already measured qubits. 

\begin{example}
$$t = (\{1,2,3\},\{1\},\{3\}, X_3^{s_1+s_2}M_2^{\alpha}X_2^{s_1}M_1^{0}E_{1,2}E_{2,3}E_{1,3})$$
Note that the commands are  read from right to left. This pattern is implementing an $Z$-rotation $R_z(-\alpha)$ from qubit $1$ to qubit $3$ (see section \ref{sec:mp-qc} for the definition of $R_z(\alpha)$).
\end{example}

Another kind of command, the \emph{dependant measurements} exist in the measurement calculus. They can be defined as a combination of measurements and dependant corrections:
$${}^\tau [M_i^\alpha]^\sigma := M_i^\alpha X^\sigma_iZ_i^\tau = M_i^{(-1)^s\alpha + \tau\pi}$$

Corrections of the form $X^{s_j+\ldots +s_k}$ clearly illustrates the hybrid nature of measurement-based  quantum computing: a classical control is collecting classical outcomes of previous measurements and computes the sum for deciding what the next quantum operation is.

Let $t_1=(V_1,I_1,O_1, A_1)$ and $t_2=(V_2,I_2,O_2, A_2)$ be two measurement patterns.
The sequential composition $t_2\circ t_1$   is  defined as $$t_2\circ t_1 := (V_1\cup V_2, I_1, O_2, A_2A_1)$$ where we assume, up to a relabelling of the qubits in $V_2\setminus O_1$ that  $V_1\cap V_2= O_1\cap I_2$. 

The parallel (or tensor) composition  $t_1\otimes t_2$ is defined as   $$t_1 \otimes t_2 := (V_1\cup V_2, I_1\cup I_2, O_1\cup O_2, A_2A_1)$$ where we assume, up to a relabelling of the qubits in $V_2$ that  $V_1\cap V_2=\emptyset$.

The size of a command is the number of qubits affected by it. Notice that for any finite sums $\sigma$ and $\tau$ of classical outcomes, $X_i^\sigma$, $Z_i^\tau$ and ${}^\tau [M_i^\alpha]^\sigma$ are of size $1$. The \emph{size} of a measurement pattern is the total size of all its commands.  The classical dependency between one-qubit operations is an important ingredient of   the depth of a measurement pattern. 
Indeed, a correction of the form $X_i^{s_j}$ has to be applied after the measurement of the qubit $j$.

The depth of a measurement pattern is longest path of dependant commands:

\begin{definition}[Quantum Depth]
For a given patterm $t=(V,I,O,A)$, its quantum depth $\depth(t)$ is defined as the longest sub-sequence $(p_x)$ of $A$ s.t. for any $x$, $dom(p_x)\cap dom(p_{x+1})\neq \emptyset$, where  $dom(E_{i,j}):=\{i,j\}$, $dom(X_i^{\sigma})= dom(Z_i^{\sigma}):=\{i\}\cup\{j\text{ s.t. $s_j$ appears in $\sigma$}\}$, $dom(M_i^\alpha):=\{i\}$ and $dom({}^\tau [M_i^\alpha]^\sigma):=\{i\}\cup\{j\text{ s.t. $s_j$ appears in $\sigma$ or in $\tau$}\}$
\end{definition}

\begin{example}
The measurement pattern  $t = (\{1,2,3\},\{1\},\{3\}, X_3^{s_1+s_2}M_2^{\alpha}X_2^{s_1}M_1^{0}$ $E_{1,2}E_{2,3}E_{1,3})$ is of size $10$ and depth $6$ ($X_3^{s_1+s_2}M_2^{\alpha}X_2^{s_1}M_1^{0}E_{1,2}E_{1,3}$ is a dependant sub-sequence of size $6$).
\end{example}

Notice that the quantum depth does not take into account the depth of the classical side-processing coming from computation of the classical sums. As a consequence, the quantum depth of a measurement pattern is based on the assumption that the classical computation is free, which can be motivated by the fact that the physical implementation of the quantum part of the computation is much more challenging than the classical part which can be considered at first approximation as free. This assumption is discussed in details in section \ref{sec:assumption}.

For any measurement patterns $t_1$ and $t_2$, $\size(t_1\otimes t_2)=\size(t_2\circ t_1)= \size(t_1)+\size(t_2)$. Moreover, $\depth(t_1\otimes t_2)=max(\depth(t_1),\depth(t_2))$ and $\depth(t_2\circ t_1)\le \depth(t_1)+\depth(t_2)$.

Since ${}^\tau [M_i^\alpha]^\sigma=  M_i^\alpha X^\sigma_iZ_i^\tau$, any measurement pattern $t$ can be rewritten into a measurement pattern $t'$ without dependant measurements such that $\size(t')\le 3.\size(t)$ and $\depth(t')=2.\depth(t)$. As a consequence, w.l.o.g, we consider in the rest of the paper only measurement patterns without dependant measurements.
 
In the following, we define some classes of complexity for measurement patterns. We consider only \emph{uniform families}, whose description can be generated by a log-space Turing machine. Moreover we consider a fixed basis of measurement angles i.e. $0,\pi$ and an irrational multiple of $\pi$.

\begin{definition}
$\MNC(d(n))$ contains decision problems computed exactly by
uniform families of measurement patterns of input size $n$, depth $O(d(n))$,
polynomial size, and over a fixed basis.
Let $\MNC^k = \MNC(\log^k n)$ and  
$\BMNC^k$ contain decision problems computed with two-sided, polynomially small error.
\end{definition}

\section{Measurement patterns and quantum circuits}\label{sec:mp-qc}

The comparison of the computational power of quantum circuits and measurement patterns has been extensively studied \cite{RBB03,BK07}. Indeed, since the introduction of the one-way model, the advantage of the one-way model in terms of depth complexity has been pointed out on some examples. 
 In this section, we review the main results and state them in terms of complexity classes.

A quantum circuit is a sequence of quantum gates. These gates are acting on three kinds of qubits: \emph{input}, \emph{output} and \emph{ancilla} qubits. Input and output qubits may overlap\footnote{This definition slightly generalises the usual definition of quantum circuit where input and output qubits are the same.}.   A quantum circuit is a quadruplet $C=(V,I,O,G)$, where $V$ is the set of all the qubits (input, output and ancilla qubits), $I,O\subset V$ are sets of input and output qubits. $G$ is a sequence of gates. The size of a gate is the number of affected qubits, the size of a circuit is the total size of all its gates. The depth  is longest path of dependant gates. 
A gate is a unitary operation applied on a bounded number of qubits. We consider the following one- and two-qubit gates: $H$, $R_z(\alpha)$ and $\Lambda Z$ (see section 2 for a definition of $\Lambda Z$):

$$H:=\frac1{\sqrt 2}\left(\begin{array}{cc}1&1\\1&-1\end{array}\right)~~~R_z(\alpha):=\left(\begin{array}{cc}1&0\\0&e^{i\alpha}\end{array}\right)$$

$\{H,R_z(\alpha_0),\Lambda Z\}$ with $\alpha_0$ an irrational multiple of $\pi$ is a universal family that can approximate any quantum gate with good precision \cite{ADH97}.

Running a quantum circuit consists in  initialising the non input qubits in the $\ket 0$ state, then applying sequence of gates, and finally measuring the non output qubits in the standard basis. 

We consider the classes of complexity $\QNC$ and $\QNC^k$ for quantum circuits \cite{QNC}:   
 $\QNC(d(n))$ contains decision problems computed exactly by
uniform families of quantum circuits of input size $n$, depth $O(d(n))$,
polynomial size, and over a fixed basis.
Let $\QNC^k = \QNC(\log^k n)$, and  
$\BQNC^k$ contain decision problems computed  with two-sided, polynomially small error.

~

Measurement patterns can be translated into quantum circuits and vice versa.

\begin{lemma}[\cite{RBB03}]\label{lem:qc-to-mp}
Any quantum circuit $C$ can be simulated by a measurement pattern $t$  of size $O(\size(C))$ and depth $O(\depth(C))$.
\end{lemma}

\begin{lemma}[\cite{BK07}, Lemma 7.9]\label{thm:mp-to-qc}
Any measurement pattern $t$ can be simulated by a quantum circuit  $C$  of size $O(\size(t)^3)$ and depth $O(\depth(t)\log(\size(t)))$.
\end{lemma}

These lemmas lead to the following inclusions of complexity classes:

\begin{theorem}for any $k\in \mathbb N$, 
$$\QNC^k\subseteq \MNC^k \subseteq \QNC^{k+1}$$
$$\BQNC^k\subseteq \BMNC^k \subseteq \BQNC^{k+1}$$
\end{theorem}

\begin{proof}
Lemma \ref{lem:qc-to-mp} implies $\QNC^k\subseteq \MNC^k$. Moreover, for any measurement pattern $t$ of input size $n$ and polynomial size, $\log(\size(t))=O(\log(n))$. So, according to Lemma \ref{lem:qc-to-mp}, $\MNC^k\subseteq \QNC^{k+1}$. $\hfill \Box$\end{proof}

There is potentially a logarithmic depth separation between quantum circuits and measurement patterns, and indeed such a separation has been pointed out for the computation of the PARITY:

$$\Upar n= \ket {x_1,\ldots ,x_n} \mapsto \ket{x_1, \ldots, x_{n-1}, \bigoplus_{i=1\ldots n}x_i}$$

$U_{parity}$ is a so called Clifford operation, a class of operations that can be computed in constant depth using measurements patterns \cite{RBB03}. On the other hand, the depth of  a  quantum circuit  for computing $\Upar n$ i of depth $\Omega(\log(n))$ \cite{BK07}. It implies that $\QNC^0\neq \MNC^0$. Such a separation is an open question for $k>0$.

 In the next section, we characterise the computational power of the measurement patterns using a reduction to quantum circuits with unbounded fan-out.

\section{Measurement patterns and quantum circuits with unbounded fan-out}\label{sec:fan-out}

Measurement patterns and quantum circuits does not have the same computational power. In this section we compare the computational power of the measurement patterns with a stronger version of the quantum circuits, the quantum circuits with unbounded fan-out. This model of quantum circuits was introduced by H\o yer and Spalek \cite{HS05}. The original motivation for introducing such a model of quantum circuits with unbounded fan-out circuits is that in addition to parallelising  operations acting on distinct qubits, commuting operations can also,  in certain circumstances,  be applied simultaneously. 
See \cite{HS05} for details on quantum fan-out circuits. A quantum circuit with unbound fan-out is a circuit with the usual gates and also fan-out gates which have unbounded input/output size:

\begin{definition}[fan-out gate]
Fan-out gate maps $$\Ufan n = \ket {y_1,\ldots , y_{n-1},x} \mapsto \ket{y_1\oplus x, \ldots, y_{n-1} \oplus x, x}$$ 
\end{definition} 

The depth and size are defined like for quantum circuits. Notice that a fan-out gate acting on $n$ qubits has size $n$. We consider the classes of complexity $\QNCwf$ and $\QNCwf^k$ for quantum circuits \cite{HS05}:   
$\QNCwf(d(n))$ contains decision problems computed exactly by
uniform families of quantum circuits with  unbounded fan-out of depth $O(d(n))$,
polynomial size, and over a fixed basis.
Let $\QNCwf^k = \QNCwf(\log^k n)$, and $\BQNCwf^k$  contain decision problems computed  with two-sided, polynomially small error.

\begin{lemma}\label{lemma:parityfan-out}
There exists a fan-out circuit of depth $3$ which implements the parity gate $$\Upar n = \ket {x_1,\ldots ,x_n} \mapsto \ket{x_1, \ldots, x_{n-1}, \bigoplus_{i=1\ldots n}x_i}$$
\end{lemma}

\begin{proof} $\Upar n = H^{\otimes n} \circ \Ufan n \circ H^{\otimes n}$ \hfill$\Box$\end{proof}

Quantum fan-out circuit can be used to parallelising commuting operations if a basis change making these operators diagonal can be implemented efficiently:

\begin{lemma}
\label{lem:commuting}
{\cite[Theorem~1.3.19]{hj:matrix-book}}
For every set of pairwise commuting unitary gates, there exists an
orthogonal basis in which all the gates are diagonal.
\end{lemma}

\begin{theorem}
\label{th:parallel}
{\cite{moore:parallel-qc,ghmp:qacc}} 
Let $\{U_i \}_{i=1}^n$ be pairwise commuting gates on $k$ qubits.  Gate
$U_i$ is controlled by qubit $x_i$.  Let $T$ be a gate changing the
basis according to Lemma~\ref{lem:commuting}.  There exists a quantum
circuit with fan-out computing $U = \prod_{i=1}^n \Lambda_{x_i}U_i$ having
depth ${\mathrm{max}_{i=1}^n \depth({U_i}) + 4\cdot \depth{(T)} + 2}$,
size $\sum_{i=1}^n \size{(U_i)} + (2n + 2) \cdot \size{(T)} + 2 n$, 
and using $(n-1)k$ ancillas.
\end{theorem}

In the following we prove that  the quantum depth of the measurement patterns and the depth of the quantum circuits with unbounded fan-out coincide. First, we prove that measurement patterns are at least as powerful as quantum circuits with unbounded fan-out:

\begin{lemma}\label{lem:mp-qfoc}
Any quantum circuit with unbounded fan-out $C$ can be simulated by a measurement pattern of depth $O(\depth(C))$ and size $O(\size(C)^3)$. 
\end{lemma}

\begin{proof} Each gate of $C$ can be simulated by a measurement pattern. Indeed $H$, $R_z(\alpha_0)$, $\Lambda Z$ can be simulated by constant depth, constant size measurement patterns: 
\begin{eqnarray*}t_H&=&(\{1,2\}, \{1\}, \{2\}, X^{s_1}_2M_1^0E_{1,2})\\t_{R_z(\alpha)}&=&t_H\circ(\{0,1\},\{0\}, \{1\}, X^{s_1}_2M_1^\alpha E_{1,2})\\t_{\Lambda Z}&=&(\{1,2\},\{1,2\},\{1,2\}, E_{1,2})\end{eqnarray*}  Since $\Ufan n$ is equal to $H^{\otimes n} \circ \Upar n \circ H^{\otimes n}$ and since it  exists a measurement pattern for $\Upar n$ of depth $O(1)$ and size $O(n^3)$ (see section \ref{sec:mp-qc}), $\Ufan n$ can be simulated by a measurement pattern of constant depth and $O(n^3)$ size. By sequential and parallel compositions, $C$ is simulated by a measurement pattern $t$ of size $O(\size(C)^3)$ and depth $O(\depth(C))$.$\hfill \Box$
\end{proof}

Moreover, quantum circuits with unbounded circuits are at least as powerful as measurement patterns:

\begin{lemma}\label{lem:qfoc-mp}
Any measurement pattern $t$ can be implemented by a quantum circuit with unbounded fan-out of depth $O(\depth(t))$ and size $O(\size(t)^2)$.
\end{lemma}

\begin{proof} 
For a given measurement pattern $t=(V,I,O,A)$, the sequence of commands $A$  can be rewritten into $k=\depth(t)$ layers $A^{(i)}$ of depth $1$ such that $A=A^{(k)}\ldots A^{(1)}$.
In the following we show that each of these layers can be translated into a constant depth piece of quantum fan-out circuit. 
Given a layer $A^{(i)}$, since $A^{(i)}$ is of depth $1$, it implies that each operation is acting on distinct qubits. Thus, up to some commutations, we assume w.l.o.g. that the commands of $A^{(i)}$ are performed in the following order: the entangling operations first, followed by measurements, $Z$-corrections and finally $X$-corrections. \begin{itemize}
\item  The sub-sequence of $A^{(i)}$ composed of entangling operations is translated into a quantum circuit composed of $\Lambda Z$. The depth of this circuit is one since the entangling operations are acting on distinct qubits.
\item   The sub-sequence of $A^{(i)}$ composed of measurements is translated into a quantum circuit where each measurement $M_j^{\alpha_j}$ is replaced by a $H_j R_z(-\alpha_j)$. The depth of this circuit is $2$.
\item The sub-sequence of $A^{(i)}$ composed of $Z$-corrections is translated to a quantum circuit where each $Z_j^\sigma$ is replaced by $\prod_{s_k\in \sigma}\Lambda_k Z_j$. This piece of circuit is not of constant depth, however, all the unitary transformations are diagonal, thus according to the Theorem \ref{th:parallel}, this piece of circuit can be simulated by  quantum circuit with unbounded fan out of constant depth and size $O(\size(t))$.
\item  Similarly, the sub-sequence of $A^{(i)}$ composed of $X$-corrections is translated into a quantum circuit with unbounded fan out of constant depth and size $O(\size(t))$. In this case, since $X$ is not diagonal, $H$ is used as basis change in Theorem \ref{th:parallel}.
\end{itemize}

Thus each layer $A^{(i)}$ is translated to a piece of circuit of constant depth and size $O(s.\size(t))$, where $s$ is the size of $A^{(i)}$. As a consequence the whole pattern is translated into a quantum circuit with unbounded fan-out of detph $O(\depth(t))$ and size $O(\size(t)^2)$. 
$\hfill \Box$ 
\end{proof}

The combination of Lemmas \ref{lem:mp-qfoc} and \ref{lem:qfoc-mp} leads us to the main result of this paper. It implies that measurement patterns and quantum circuits with unbounded fan-out have the same computational power:
\begin{theorem}\label{thm:same-power}
For any $k\in \mathbb N$, $$\MNC^k = \QNCwf^k$$
$$\BMNC^k = \BQNCwf^k$$
\end{theorem}

\section{Generalisation to classically controlled quantum computation}\label{sec:gen}
In this section a general scheme of classically controlled quantum computation is considered. One-way quantum computation is a special instance of this scheme as well as the teleportation based model \cite{N01}, the state transfer model \cite{Per05a} and the ancilla-driven quantum computation with twisted graph states \cite{KOBAA09}.

A classically controlled scheme, generalization of the measurement calculus, is characterized by a  quadruplet $(\mathcal I,\mathcal U,\mathcal O,\mathcal C)$ where $\mathcal I$ is a set of quantum states (for initialising ancillary qubits); $\mathcal U$ is a set of  unitary transformations; $\mathcal O$ is a set of measurements  with classical outcomes in $\{0,1\}$;  $\mathcal C$ is a set of  corrections that are classically controlled by a sum modulo two of measurement outcomes. 

For instance, the one-way model  is a $(\{\ket +\},\{\Lambda Z\}, \{\{\ket{+_\alpha},\ket{-_\alpha}\}, {\alpha\in [0,2\pi)}\},$ $ \{Z,X\})$-scheme; 
the state transfer model \cite{Per05a} is a $(\emptyset, \emptyset,\{X\otimes Z, Z, (X-Y)/\sqrt 2\}, \{Z,X\})$-scheme; the quantum circuit model is a $(\{\ket 0\}, \{H,T,\Lambda Z\},$ $ \{\{\ket 0,\ket 1\}\}, \emptyset)$-scheme; and the ancilla-driven model \cite{KOBAA09} is a $(\{\ket +\},$ $ \{(H\otimes H) \circ \Lambda Z\}, \{\{\ket{+_\alpha},\ket{-_\alpha}\}, {\alpha\in [0,2\pi)}\}, \{Z,X\})$-scheme.

Size and depth of a classically controlled pattern are defined as for the measurement patterns, where every primitive operation has a constant  depth and its size is the number of qubits affected by it.

The classes $\QNCwf^k$ can be extended to unitary transformations, quantum states and quantum measurements as follows: a unitary $U$ is in  $\UQNCwf^k$ if $U$ can be implemented by a quantum circuit with unbounded fan-out of depth $k$; a quantum state $\ket \phi$ is in  $\IQNCwf^k$ if there exists $U\in \UQNCwf^k$ such that $\ket \phi =U\ket 0$; finally a linear map $O$ is in  $\OQNCwf^k$ if there exists $U\in \UQNCwf^k$ such that $UOU^\dagger$ is diagonal. If $O\in \OQNCwf^k$ is an observable (self adjoint) then $O$ is describing a measurement that can be implemented in depth $O(k)$ by a circuit with unbounded fan-out: the measurement according to $O$ is transformed into a measurement in the standard basis thanks to the unitary $U\in \OQNCwf^k$ such that $UOU^\dagger$ is diagonal. 

We show that among all the classically controlled schemes which commands can be implemented in constant depth, the measurement calculus is optimal in term of depth: 

\begin{theorem}\label{thm:gen} Given a  classically controlled scheme $(\mathcal I,\mathcal U,\mathcal O,\mathcal C)$, if $\mathcal I\subseteq \IQNCwf^0$, $\mathcal U\subseteq \UQNCwf^0$, $\mathcal O\subseteq \OQNCwf^0$, $\mathcal C\subseteq \UQNCwf^0\cap \OQNCwf^0$, then any pattern $P$ of that scheme can be implemented by  a measurement pattern $t$ of depth $O(\depth(P))$.
\end{theorem}

\begin{proof}
We prove that any classically controlled pattern $P$ can be implemented by a quantum circuit of unbounded fan-out of depth $O(\depth(P))$, which can then be implemented by a measurement pattern $t$ of depth $O(\depth(P))$ according to lemma \ref{lem:qfoc-mp}.

The proof generalizes the proof of lemma \ref{lem:qfoc-mp}. $P$ is composed of $\depth(P)$ layers $\{P^{(i)}\}_{1\le i\le depth(P)}$. Notice that for each layer, the operations are acting on distinct qubits and then can be reorganised into subsequences of each type (intialisation, unitary, measurement and correction). Each of these subsequences can be implemented in constant depth:
\begin{itemize}
\item The subsequences of $P^{(i)}$ composed of initialisations, unitaries and measurmeents  can be implemented in constant depth since $\mathcal I\subseteq \IQNCwf^0$, $\mathcal U\subseteq \UQNCwf^0$, and $\mathcal O\subseteq \OQNCwf^0$; 
\item 
The subsequence of $P^{(i)}$ composed of  corrections can also be implemented in constant depth. Indeed, since every correction of the subsequence are acting on distinct qubit, all these corrections are commuting. Thus we can apply theorem \ref{th:parallel}, leading to a quantum circuit with unbounded fanout of constant depth since each $C\in \mathcal C$ is of constant depth and the gate $T$ changing the basis is of constant depth as well since $\mathcal C\subseteq \OQNCwf^0$.
\end{itemize}
Thus each layer is implemented by a constant depth piece of circuit, so $P$ is translated into a unbounded fan-out circuit of depth $\depth(P)$.\hfill$\Box$
\end{proof}

Notice that the one-way, the teleportation-based, the state-transfer-based, and the ancilla-driven models are all classically controlled models for which theorem \ref{thm:gen} applies. As a corollary, it exists a depth-preserving translation from the ancilla-driven model to the one-way model, answering an open question stated in \cite{KOBAA09}.

\section{Applications}\label{sec:app}

The complexity classes of the quantum circuits with unbounded circuits
have been studied and compared to other classes. Thanks to Theorem
\ref{thm:same-power}, all known complexity results  about quantum
circuits with unbounded fan-out can be applied to the measurement
patterns. Among them, it is known that the quantum Fourier transform
(QFT) is in $\BQNCwf^0$ \cite{HS05}, so QFT is in $\BMNC^0$. Whereas
the one-way model have mainly been introduced as a promising model of
physical implementation, it turns out that this model is very
powerful: QFT can be approximated in constant depth in the one-way
model. This result confirms that the semi-classical quantum Fourier
transform  can be done efficiently \cite{Gri96}.

Moreover, factorisation is in $\RP[\BQNCwf^0]=\RP[\BMNC^0]$
\cite{HS05}, thus the factorisation can be approximated efficiently on
a probabilistic machine which has access to a constant depth one-way
quantum computer.

\section{A weaker assumption}\label{sec:assumption}

Parallelisation in the quantum circuit model is based on the following assumption: 
(\emph{i}) gates acting on distinct qubits can be performed simultaneously.  
Measurement patterns and quantum circuits with unbounded fan-out and have an extra power compared to quantum circuits because 
they are based on stronger assumptions. The additional assumption for quantum circuits with unbounded fan-out is: 
(\emph{ii}) commuting operations can be done simultaneously.\footnote{In fact assumption $(ii)$ implies assumption $(i)$} 
Whereas the additional assumption for measurement patterns is: 
 (\emph{iii}) classical part of any measurement pattern can be done in constant depth. 

In this section, we investigate assumption $(iii)$, and we compare assumptions $(ii)$ and $(iii)$ which lead to the same extra power. We show that these two assumptions can be related, and that assumption $(iii)$ is weaker than $(ii)$.

The classical part of a measurement pattern is reduced to the computation of sum modulo $2$ (i.e., parity) \cite{AB09} in the dependant correction commands of the form $X_i^{s_i+\ldots +s_j}$.  Thus, assumption $(iii)$ can be rephrased as: \emph{boolean unbounded fan-in parity gates can be done in constant depth}.

\begin{lemma}
 If boolean unbounded fan-in parity gates can be done in constant depth, any measurement pattern $t$  can be done in $\depth(t)$  quantum layers of constant-depth interspersed by constant depth classical layers. 
 \end{lemma}

Notice that without assumption $(iii)$, the classical parity gate on $n$ bits can be  computed in depth $O(\log (n))$. Thus,

\begin{lemma}
Any measurement pattern $t$ can be done in $\depth(t)$ constant-depth quantum layers interspersed by $O(\log(\size(t)))$-depth classical layers.  
\end{lemma}

Now we show that assumption $(iii)$ is weaker than assumption $(ii)$. Indeed, assumption $(ii)$ is that commuting operations can be done simultaneously, which implies that classical\footnote{$U$ is classical if $U$ maps basis states to basis states in the computational basis.} commuting  operations can be done simultaneously. So, the classical parity can be implemented in constant depth using classical control-Not which are commuting since target and controlled bits are not overlapping. 

In other words, whereas the assumption associated to the quantum circuits with unbounded fan-out is that commuting operations can be done in parallel, the classical version of this assumption is enough for the measurement patterns.

\section{Conclusion}

In this paper, we have shown that the measurement patterns has an
equivalent computational power to the quantum circuits enhanced with
unbounded fanout. This characterises the power of the one-way model
for parallelisation of algorithms and demonstrates that a number of
quantum algorithms can be achieved in this model with a modest quantum
depth.
The equivalence of these two models is at first sight surprising. The
parallelisation structure of the one-way model would appear to be
radically different to the quantum circuit model. Nevertheless, our
results indicate a close underlying connection.

There remain a number of open questions. Can the separation
$\QNC^0\neq \MNC^0$ (proved by Broadbent and Kashefi with the parity
problem) be proved for any $k$: $\QNC^k\neq \MNC^k$? Or at least for
$k=1$. Moreover, given the examples above, to what extent can quantum depth
be minimised for generic quantum algorithms? Recently, it has been shown that non-trivial quantum algorithms can be
implemented with a single round of simultaneously applied commuting
gates \cite{SB08}. Moreover,  in \cite{jozsa} Richard
Josza made the following conjecture:

``Any polynomial time quantum algorithm can be implemented with only
$O(\log n)$ quantum layers interspersed with polynomial time classical
computations. ''

Our results may have an implication for fault tolerant quantum
computation. Thresholds for fault tolerant computation are typically
derived under the assumption of a polynomial computational depth. How
could such fault tolerant models be relaxed if only constant or
logarithmic quantum depth would suffice? We have focussed on classically controlled quantum computation where the classical control consists sums modulo two in our study. These results applies to one-way quantum computation and various measurement-based models, like the teleportation-based and the ancilla-driven models. But, other
variants of measurement-based quantum computation with a different
dependency structure \cite{eisertgross} could have a quite different
characterisation. We hope that this work motivates further study of this rich area.

\end{document}